\documentclass[twocolumn,secnumarabic,amssymb, nobibnotes, aps, prl, superscriptaddress, nobalancelastpage,longbibliography,nofootinbib]{revtex4-1}

\usepackage{graphicx}
\usepackage{amsfonts}
\usepackage{amsmath,amssymb}
\usepackage{bm}
\usepackage{xcolor}
\usepackage{float}
\usepackage{dcolumn}
\usepackage{cancel}

\usepackage{xr-hyper}
\usepackage{hyperref}
\hypersetup{breaklinks=true,colorlinks=true,linkcolor=blue,citecolor=blue,filecolor=magenta,urlcolor=cyan}

\usepackage[all]{hypcap}

\newcommand{\Apv}{A_{\rm PV}}
\newcommand{\aD}{\alpha_{\rm D}}
\newcommand{\skin}{R_{\rm skin}}

\newcommand{\Pb}{$^{208}$Pb}
\newcommand{\Ca}{$^{48}$Ca}

\makeatletter
\def\@bibdataout@aps{%
\immediate\write\@bibdataout{%
@CONTROL{%
apsrev41Control%
\longbibliography@sw{%
    ,author="08",editor="1",pages="1",title="0",year="1"%
    }{%
    ,author="08",editor="1",pages="1",title="",year="1"%
    }%
  }%
}%
\if@filesw \immediate \write \@auxout {\string \citation {apsrev41Control}}\fi
}
\makeatother

\begin{document}

\title{Combined theoretical analysis of the parity-violating asymmetry for {\Ca}  and {\Pb} }

\author{Paul-Gerhard Reinhard}
\affiliation{Institut für Theoretische Physik, Universität Erlangen, Erlangen, Germany}
\email{paul-gerhard.reinhard@fau.de}

\author{Xavier Roca-Maza}
\affiliation{Dipartimento di Fisica ``Aldo Pontremoli'', Universit\`a degli Studi di Milano, 20133 Milano, Italy and INFN, Sezione di Milano, 20133 Milano, Italy}
\email{xavier.roca.maza@mi.infn.it}

\author{Witold Nazarewicz}
\affiliation{Facility for Rare Isotope Beams and Department of Physics and Astronomy, Michigan State University, East Lansing, Michigan 48824, USA}
\email{witek@frib.msu.edu}

\date{\today}
\begin{abstract}
The recent experimental determination of the parity violating asymmetry $\Apv$ in ${}^{48}$Ca and ${}^{208}$Pb at  Jefferson Lab is important for our understanding on how neutrons and protons arrange themselves inside the atomic nucleus. To better understand the impact of these measurements, we present a rigorous theoretical investigation of $\Apv$ in {\Ca} and {\Pb} and assess the  associated uncertainties. We complement our study by inspecting  the static electric dipole  polarizability in these nuclei. The analysis is carried out within  nuclear energy density functional theory with quantified input.  We conclude that the simultaneous accurate description of $\Apv$  in ${}^{48}$Ca and ${}^{208}$Pb cannot be achieved by our models that  accommodate  a pool of global nuclear properties, such as masses and charge radii, throughout the  nuclear chart,  and describe - -within one standard deviation-- the experimental dipole polarizabilities $\aD$ in these nuclei.
\end{abstract}
\maketitle

\paragraph{Introduction.} Polarized elastic electron scattering and polarized proton scattering have been recently used at Jefferson Lab \cite{PREX-2,CREX} and  RCNP in Osaka \cite{Tamii2011,Birkhan2017} to measure, respectively,   $\Apv$ and  $\aD$ in {\Ca} and {\Pb}.  These nuclei are the two stable doubly-magic systems that have substantial neutron-to-proton  asymmetry measured in terms of the neutron excess $N-Z$. Large neutron excess increases the neutron skin thickness, decreases $\Apv$, and increases $\aD$, and  many aspects of theoretical description get simplified in  doubly-magic nuclei, which makes them particularly attractive for theory. Moreover, to connect properties of the atomic nucleus to the nuclear matter  equation of state (EoS), it is preferable to study heavy  systems such as  {\Pb} whose properties are dominated by volume effects.

Since the  electron scattering, governed by the electroweak force,  is relatively well understood, it promises a clear interpretation of results. 
In this respect, it should be noted  that in order to extract information on the neutron skin thickness $\skin$ and   the symmetry energy parameters $J$ and $L$
of the EoS
from the observed  $\Apv$ at a single
kinematic condition, nuclear models must be used. The  dependence of the results on a nuclear model $\cal M$ enters through (i) the description of the parity-violating response \cite{Donnelly1990} and (ii) the nuclear model of electroweak charge distribution of the atomic nucleus. This model dependence results in  uncertainties which need to be considered when carrying out the extraction $\Apv \xrightarrow{\cal M} \skin, J, L$
\cite{Reinhard2021}. In the case of $\aD$, the model dependence in the analysis stems from
 distorted wave impulse approximation analysis of proton scattering data, including assumed optical potential model \cite{Tamii2009}, and
 possible contaminations of the $E$1 nuclear response from (i) other nuclear multipolarities and (ii) quasi-deuteron excitations.

The  PREX-2 \cite{PREX-2} result  has stimulated a number of studies with often contradictory results  on the impact of  $\Apv$ on  various nuclear observables and astrophysical data. For example, in Refs.~\cite{PREX-2,Reed2021}, a particular set of covariant energy density functionals (EDFs) was used to infer information on $\skin$, $J$, and $L$ as well as on some neutron star properties. Using the same family of EDFs,  Ref.~\cite{Piekarewicz2021} analysed implications of the PREX-2 on $\aD$ and concluded that  there exists a tension between the value of $\skin$ reported by  PREX-2 and measured value of $\aD$. On the other hand, the reaction cross-sections for proton and alpha  scattering  \cite{Wakasa2021,Shingo2021,Matsuzaki2021} were found to be consistent with the large value of $\skin$ deduced by PREX-2.  In Ref.~\cite{Reinhard2021},  $\Apv$ was analyzed by  taking special care of model uncertainties  and correlations with other observables such as $\aD$. According to this work, the significant 1-$\sigma$ uncertainty of PREX-2 value of $\Apv$ precludes the use of this observable as a constraint on the isovector sector of current EDFs. Other studies \cite{Li2021b, Essick2021,Essick2021a, Most2021, Newton2021, Biswas2021, Lattimer2021,Sammarruca2021,Zhang2021,Baishan2021,Lim2022} also found it difficult to accommodate the PREX-2 values of $\skin$ and $L$.
We note that some of these references consider the value of $\skin$ reported in Ref.~\cite{PREX-2} as a {\it measured} quantity, ignoring the aspect of the  model-dependent extraction  $\Apv \xrightarrow{\cal M} \skin$.

In this Letter, we carry out a comprehensive theoretical investigation of $\Apv$ and $\aD$ within nuclear density functional theory (DFT) \cite{Bender2003} supplemented by statistical uncertainty quantification and correlation analysis. In this way, we assess the impact  of $\Apv$ in {\Pb} and {\Ca} on EDFs developments and on the nuclear matter symmetry energy at saturation.

\paragraph{Parity violating asymmetry.} Polarized elastic electron scattering gives access to the parity violating asymmetry $\Apv$, an observable that probes the weak charge density distribution in atomic nuclei provided the electromagnetic charge density is known \cite{Donnelly1990,Horowitz2001}. Via theoretical models, $\Apv$ has been used to extract information on the neutron skin thickness and on the symmetry-energy parameters  $J$ and  $L$ (see, e.g., \cite{Roca-Maza2011,Reed2021,Reinhard2021}). For an accurate analysis of the measured $\Apv$, different contributions must be considered. In medium mass and heavy nuclei such as {\Ca} or {\Pb}, Coulomb distortions must be accounted for \cite{Horowitz1998}. Accurate nucleon electromagnetic and weak form factors are essential \cite{Horowitz2001}. A correct understanding of the beam polarization is also crucial. In this respect, the analyzing power obtained in the PREX-2 experiment is quite puzzling \cite{PREX-Bn1, Esser2018, Esser2020, PREX-Bn2, Koshchii2021}. At high incident electron-beam energies, inelastically scattered electrons from low-energy excited states or even from the giant dipole resonance of the studied target may   impact results \cite{CREX, Horowizt2014}. Effects from QED corrections to the Coulomb field felt by the incident electrons as well as radiative processes such as Bremsstrahlung have  not been estimated in this context.  For small  enough scattering angles, even atomic electrons may display some impact on $\Apv$. Finally, the currently-neglected higher-order contributions  to $\Apv$, such as magnetic effects, or two-body currents, may play some role (see also Supplemental Material (SMat) \cite{SM}). 
\nocite{Atac2021,PDG2018,Horowitz2012,Hoferichter2020,HAPPEX,Liu2007,Alexandrou2020,Reinhard2021so}

\begin{table}[htb]
\caption{\label{tab:APV-Ca} 
Parameters and results of the CREX experiment \cite{CREX}.
}
\begin{ruledtabular}
\begin{tabular}{lcc}
mean scattering angle: & $\overline{\theta}_\mathrm{Ca}$ & $4.51 \pm 0.02$ \\
transferred momentum: & $\langle Q^2\rangle$ & 
                 $0.0297 \pm 0.0002$\,GeV$^2$ \\
   & $q$ & $0.8733\pm 0$\,fm$^{-1}$\\
beam energy: & $E_\mathrm{beam}$ & $2182 \pm 0.5$ MeV \\
weak charge: & $Q_W$ & $26.0\pm 0.1$ \\
\hline
parity viol. asymmetry: & $A_\mathrm{PV}^\mathrm{(Ca)}$ & $2668 \pm 106$\,ppb\\
weak form factor at $Q^2$: & $F_W^\mathrm{(Ca)}$ & $0.1304\pm 0.0072\%$
\end{tabular}
\end{ruledtabular}
\end{table}
We show in Table\,\ref{tab:APV-Ca} the parameters and results of  CREX  \cite{CREX}. Nucleonic moments which are also needed for processing the data are given in Table\,S1 of SMat \cite{SM}. 
In this study,  we have performed calculations of $\Apv$ in {\Ca} using the same parameters/conditions as in  experiment, including the reported acceptance function.
Our calculations strictly follow those of Ref.~\cite{Reinhard2021} for $\Apv$  in {\Pb} 
based on quantified EDFs. For more details, see  SMat. 

\paragraph{Dipole polarizability.} 
The dipole polarizability $\aD$ quantifies the 
restoring force of the nucleus if an external electric dipole field tries to pull away protons from neutrons. Hence, this quantity characterizes  the isovector channel via the average symmetry potential felt by nucleons. Experimentally, $\aD$ can be deduced using real photons from the total photo-absorption cross section \cite{Ahrens1975} or, equivalently, using virtual photons in polarized proton scattering \cite{Tamii2011,Birkhan2017}. Theoretically,
$\aD$  can be computed from integrating the inverse-energy weighted dipole strength distribution \cite{Reinhard2010,Piekarewicz2012, Reinhard2013a,Roca-Maza2013_2,Hashimoto2015,  Roca-Maza2015, Bassauer2020, Goriely2020}. The results presented here are based on the latter approach by using the same EDFs employed to calculate $\Apv$. Within this framework, the product of  $\aD$ and $J$ has been shown to be linearly correlated with the neutron skin thickness or, similarly, with the $L$ parameter in neutron rich medium and heavy nuclei \cite{Roca-Maza2013_2, Roca-Maza2015}.

\paragraph{Parametrizations and observables.} We base our study on three different types of EDFs. This serves to assess the impact of the form of a functional. One EDF is of the non-relativistic standard Skyrme type, labeled SV \cite{Klupfel2009}; the second EDF, labeled RD,  is a generalized Skyrme type that contains a richer density dependence in terms of rational approximants \cite{Erler2010}; and the third one is a point-coupling relativistic mean-field EDF, labeled  PC \cite{Niksic2008}, optimized using  the same  dataset
as SV-min \cite{Nazarewicz2014}. All three functional families are optimized with respect to the same set of ground-state data, energies, charge radii, surface thickness, etc., in more than 60 semi-magic, spherical nuclei \cite{Klupfel2009}. In addition, in SMat, we explore the impact of model extensions by considering a  Skyrme parametrization SV-ext that contains  a richer density dependence than SV-min but implemented differently than in RD. 

\begin{table}[htb]
\caption{\label{tab2} Summary of the EDFs used in the present work and
  their fit observables.  All EDF parametrization use the set of ground
  state data from \cite{Klupfel2009}. 
 The various test parametrizations use additional constraining 
data on $\aD$ and $\Apv$ in {\Ca} and {\Pb} as indicated.  The parametrization SV-min$^*$ was introduced in
Ref.~\cite{Reinhard2021}.
}
\begin{ruledtabular}
\begin{tabular}{lcccc}
parametrization &  $\alpha_D(\mathrm{Ca})$ &  $A_{\rm PV}(\mathrm{Ca})$ &  $\alpha_D(\mathrm{Pb})$ &  $A_{\rm PV}(\mathrm{Pb})$ \\
\hline
 SV-min & $-$ & $-$ & $-$ & $-$
\\
 SV-APV$^2\alpha^2$ & + & +& + & +\\
 SV-APV$^1\alpha^2$ & + & + & + & $-$ \\
 SV-$\alpha^2$ & + & $-$ & + & $-$ \\
 SV-min$^*$  & $-$ & $-$ & + & + \\
\hline
 RD-min & $-$ & $-$ & $-$ & $-$\\
 RD-APV$^2\alpha^2$  & + & +& + & +\\
 RD-APV$^1\alpha^2$  & + & + & + & $-$ \\
 RD-$\alpha^2$ & + & $-$ & + & $-$ \\
\hline
 PC-min & $-$ & $-$ & $-$ & $-$
\end{tabular}
\end{ruledtabular}
\end{table} 
The basic parameterizations in each family are obtained from a fit to the given dataset. They are named SV-min \cite{Klupfel2009}, RD-min \cite{Erler2010} and PC-min \cite{Nazarewicz2014}. All of them provide high quality in the reproduction of ground-state nuclear properties. We emphasize that none of these EDFs included the data on $\Apv$ or $\aD$ in the fit. In order to assess the information content of $\Apv$ and $\aD$, we 
also develop  new parametrizations which add
the recent experimental data for $\Apv$ and/or $\aD$ in ${}^{48}$Ca and/or ${}^{208}$Pb to the fitting protocol of the SV and RD functionals as shown in Table~\ref{tab2}.
To avoid that those extended fits drive into unphysical regions, we constrain additionally three basic nuclear matter properties (NMP): incompressibility $K$; isoscalar effective mass $m^*/m$; and sum rule enhancement factor $\kappa_\mathrm{TRK}$ (isovector effective mass). These NMP are fixed such that the new  parametrizations reproduce the Giant Monopole Resonance, Giant Dipole Resonance, and Giant Quadrupole Resonance in {\Pb}  with the same quality as the original SV-min and RD-min parametrizations.

\begin{figure}[htb]
{\includegraphics[width=\linewidth]{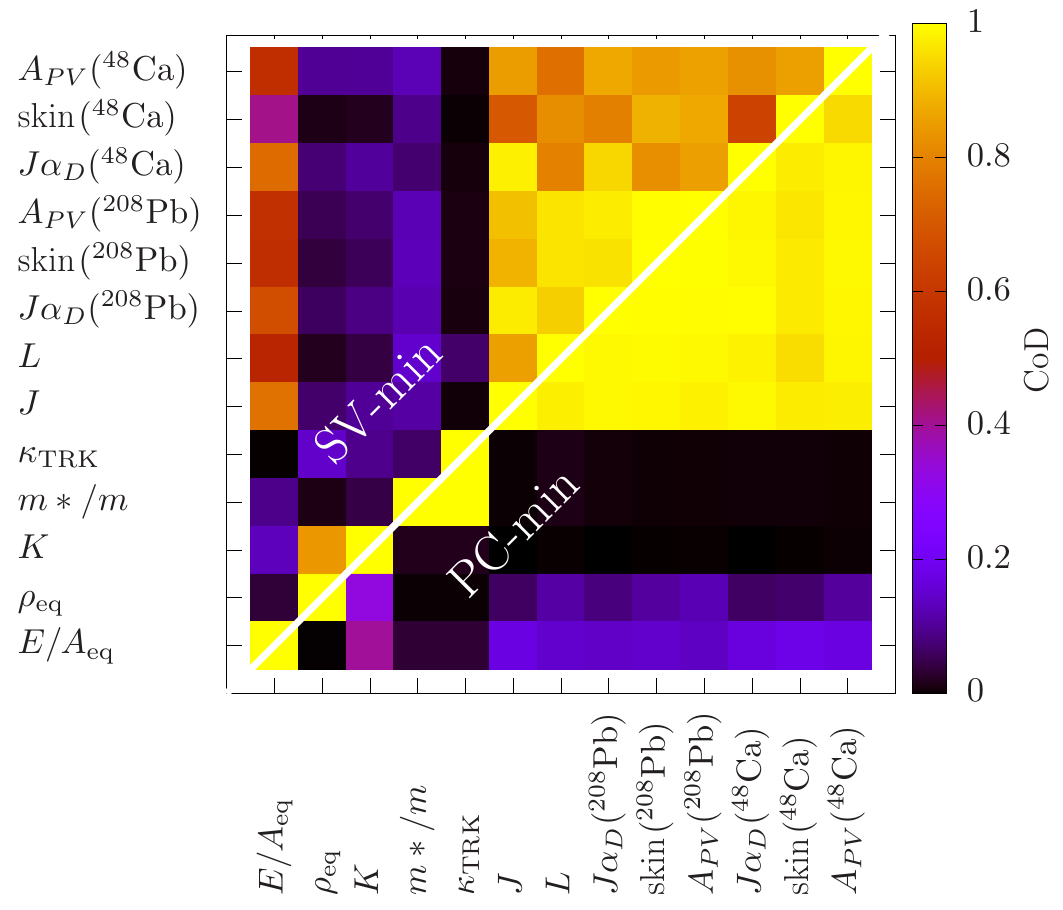}}
\caption{\label{fig:alignmatrix}
Matrix of coefficients of determination (CoD) between model parameters
and observables for SV-min (upper triangle) and PC-min (lower
triangle).
Not shown are the CoDs with spin-orbit parameters which are negligible
and with surface parameters which are small.
}
\end{figure}

\paragraph{Model parameters and nuclear matter properties}
The three functionals use to some extent different types of parameters. To make them better comparable, we express all model parameters related to bulk properties in terms of nuclear matter properties (NMP) which characterize the energy per particle ($e=E/A$) of infinite symmetric nuclear matter at zero temperature  around nuclear saturation density $\rho_\mathrm{eq}$. These can be grouped  into isoscalar and isovector NMP. The isoscalar NMP are: equilibrium energy $e_\mathrm{eq}$, equilibrium density $\rho_\mathrm{eq}$, incompressibility $K$, and isoscalar effective mass $m^*/m$. The isovector NMP are: symmetry energy $J$, slope of symmetry energy $L$, and sum rule enhancement factor $\kappa_\mathrm{TRK}$ (equivalent to isovector effective mass).
The symmetry energy slope $L$, being proportional to the pressure of pure neutron matter at saturation, is a crucial input for neutron star models.

\paragraph{Correlation analysis.} 
A  linear-regression interpretation of the $\chi^2$ fits of the parametrizations  allows to deduce uncertainties of model parameters or observables and correlations between them \cite{Reinhard2013,Dob14a,Erler2015}. A useful dimensionless measure of correlation is the coefficient of determination (CoD) between two parameters/observables \cite{Allison}.
In Fig.~\ref{fig:alignmatrix} the CoD matrix for the bulk
model parameters (those which can be expressed in terms of NMP) and the key observables of this study: $\aD$, $\Apv$, and  $\skin$ are shown. Specifically,  we show the result for two different parametrizations, SV-min  and PC-min. (The results for RD-min are very similar to those of SV-min.) Except for $J$ and $L$, other model parameters  are practically uncorrelated with the observables of interest  while
the isovector NMP $J$ and $L$  show strong correlations with $\Apv$, $\aD$, and $\skin$.
This shows that these quantities  are all isovector indicators \cite{Reinhard2010}. At least for {\Pb}, we see a 99\% correlation between $\skin$ and  $\Apv$, which means that $\Apv$ contains the information about $\skin$ for the models  considered here. As expected,
the correlations between $\skin$ and symmetry energy parameters deteriorates when going from {\Pb} to {\Ca} due to stronger surface effects in  {\Ca}.

We also see that PC-min produces stronger isovector correlations than SV-min. The reason is that the relativistic PC functional, as most other relativistic functionals, is poorly parametrized in the isovector channel which means that the isovector observables {\it must be} strongly correlated \cite{Reinhard2010}. For instance, the EDF FSUGold2 \cite{Chen2014} used in the PREX-2 analysis \cite{Reed2021}, employs only 2 isovector parameters. For Skyrme functionals, on the other hand, the parametrization of the isovector channel is  as rich as for the isoscalar channel, which yields a greater versatility at the price of requiring more isovector observables to properly determine the isovector coupling constants.

\begin{figure}[htb]
\includegraphics[width=1.0\linewidth]{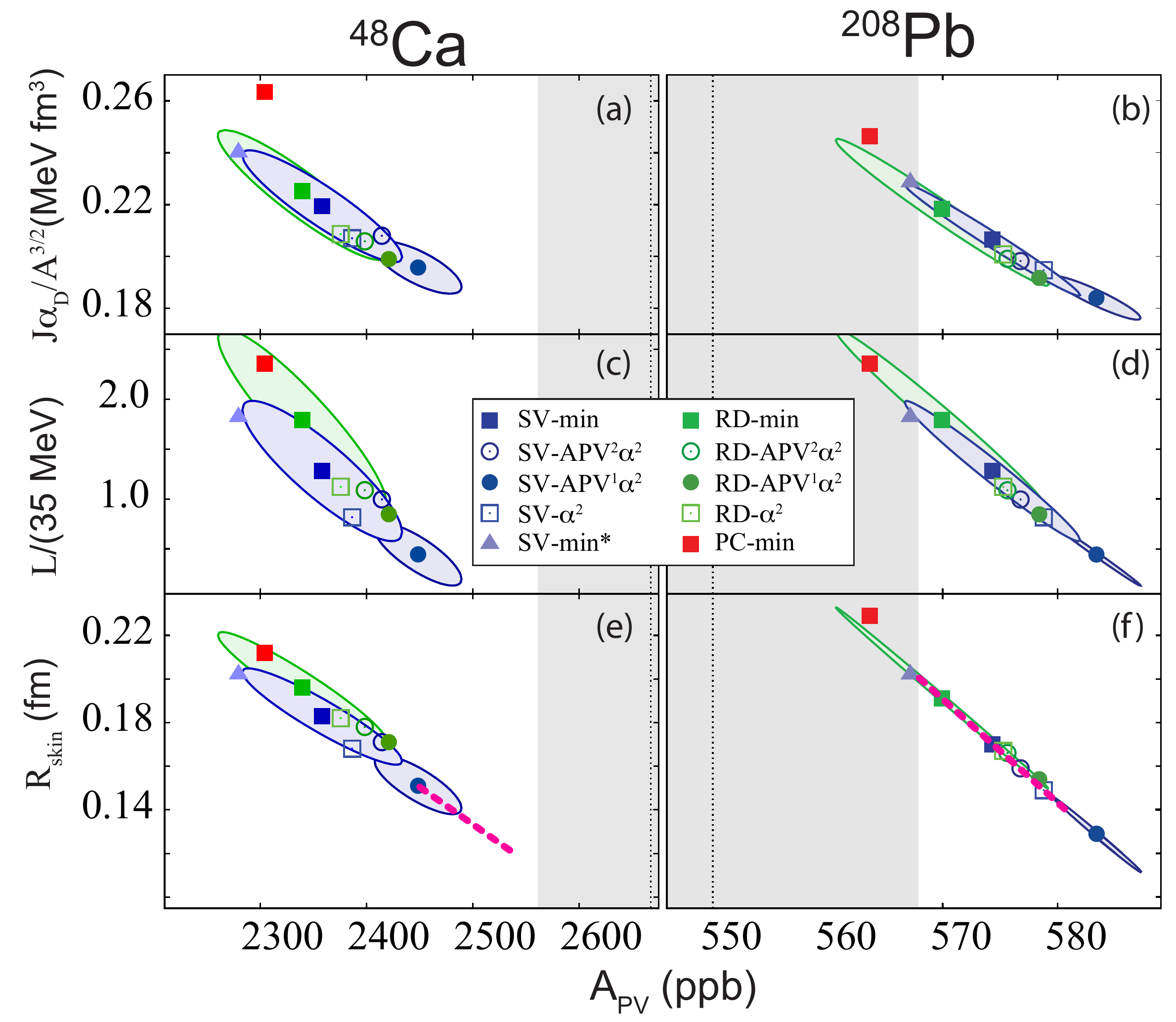}
\caption{\label{fig:APV2-trendAPV208-JL} Trend of
  $\skin$, $L$, and  $J\alpha_D$ with $\Apv$
  for {\Ca}  (left) {\Pb}  (right). Different EDFs  are distinguished by
  symbols and colors. For three parametrizations (SV-min,
  SV-APV$^1\alpha^2$, and RD-min) the error ellipsoids are indicated. Dashed magenta lines in panels (e) and (f) indicate the range of  $\skin$ predicted by ab-initio calculations of Ref.~ \cite{Hagen2016} in {\Ca} and Ref.~\cite{Baishan2021} in {\Pb} assuming the same $\Apv-\aD$ trend as in DFT calculations. The mean 
  values  of measured $\Apv$ are marked by vertical dotted lines
  and their 1-$\sigma$ errors by gray bands. 
  The values of  $\skin$ in {\Ca} for PC-min, SV-min, and SV-APV$^1\alpha^2$ are
$0.229\pm 0.027$\,fm, $0.170\pm 0.034$\,fm, and $0.129\pm 0017$\,fm, respectively. The corresponding values 
of $L$ are $82.5\pm 17.2$\,MeV, $44.8\pm 24.6$\,MeV, and $15.5\pm 11.0$\,MeV.
}
\end{figure}

\paragraph{Deducing neutron skin and isovector NMP from $\Apv$.} 
The major objective of PREX-2 and CREX experiments was to accurately measure $\Apv$ to assess the size of  $\skin$. The accompanying  theoretical analysis \cite{Reed2021} has  attempted to deduce isovector bulk NMP from the PREX-2 data using a set of relativistic functionals. How reliable is such extraction? We now discuss this question with the help  of trend analysis. Figure~\ref{fig:APV2-trendAPV208-JL} shows the trends of $J\alpha_D$, $L$, and $\skin$ with $\Apv$ for
{\Ca} and {\Pb} calculated at the experimental conditions of CREX and PREX-2. The grey regions correspond to the experimental 1-$\sigma$ error bands and the vertical dotted lines mark the mean value reported in \cite{PREX-2, CREX}. As expected, all three quantities  show a  clear trend with $\Apv$. Mind, however, that a trend alone is not conclusive as one must also inspect the variance of the prediction. This is done here by showing the error ellipsoids for three parametrizations: RD-min, SV-min, and SV-APV$^1\alpha^2$. The ellipsoids seem to align along the linear trend. Variances perpendicular to the trend are larger for {\Ca}  and very small for {\Pb}. The stronger correlations associated with ${}^{208}$Pb had already been seen in Fig.~\ref{fig:alignmatrix}.  Particularly impressive is the strong correlation between the $\skin$ and $\Apv$ in  {\Pb}   illustrated by the needle-shaped error ellipsoids for all three models shown. Only slightly weaker correlations with $\Apv$ are seen for $L$ and $J\alpha_D$. We also show in Fig.~\ref{fig:APV2-trendAPV208-JL}(e,f) the prediction of $\skin$ from the ab-initio calculations of Refs.~\cite{Hagen2016,Baishan2021}.

The correlations as such look encouraging. However, the comparison with the data on $\Apv$  is  disappointing. The theoretical predictions for  $\Apv$ tend to overestimate {\Pb}  and clearly underestimate {\Ca}. We tried to find a compromise by calibrating our models by imposing constraints on the values of $\Apv$ and $\aD$, see Table~\ref{tab2}. It is remarkable that the resulting EDFs  conform to  the linear trend. But doing so, they fail to improve the agreement with both experiments simultaneously.  Actually, most of the theoretical results shown do not overlap or barely overlap (1-$\sigma$) with the experimental data on $\Apv$. As an example,  the relativistic EDF PC-min that predicts $\Apv$ in {\Pb} consistent with experiment, spectacularly fails for {\Ca}. As discussed above, the isovector sector of PC-min is underdeveloped, and the same can be stated about the relativistic EDFs used in Ref.~\cite{Reed2021} that were used to extract the value of $\skin$ from the PREX-2 measurement.

The ab-initio calculations for {\Ca} \cite{Hagen2016} predict  $\skin$ that is smaller than the EDF models used. As seen in Fig.~\ref{fig:APV2-trendAPV208-JL}(e), this  result is more consistent with  the CREX data. Still, large deviation for the PREX data remains \cite{Baishan2021}. To make a more definite assessment of ab-initio results, their predictions for $\Apv$ would be desirable.

\begin{figure}[htb]
\includegraphics[width=0.8\linewidth]{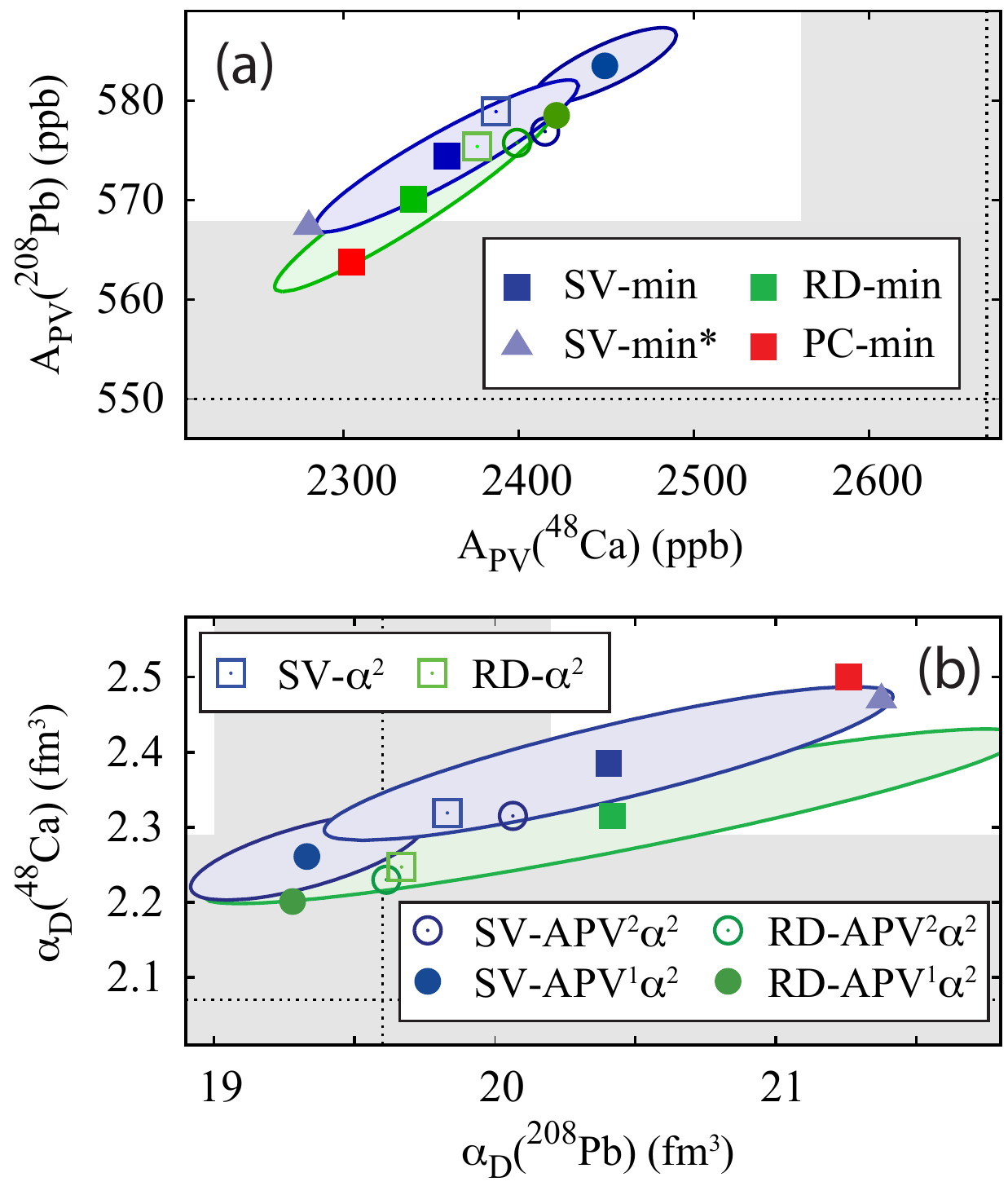}
\caption{\label{fig:APV-polariz-48vs208}
Trends of measured observables: (a)
$A_\mathrm{PV}(^{208}\mathrm{Pb})$ versus 
$A_\mathrm{PV}(^{48}\mathrm{Ca})$ and (b)
$\alpha_D(^{48}\mathrm{Ca})$ versus
$\alpha_D(^{208}\mathrm{Pb})$.
Different functionals  are distinguished by
  symbols and colors. For three parametrizations (SV-min,
SV-APV$^1\alpha^2$, RD-min), the error ellipsoids are indicated. The experimental means 
are marked by dotted lines and their errors -- by gray bands.
}
\end{figure}


\paragraph{Trends of $\Apv$ and $\aD$ in {\Ca}  versus {\Pb}.} 
The discussion of Fig.~\ref{fig:APV2-trendAPV208-JL} ends with indicating a tension between  PREX-2 and CREX  values of $\Apv$  viewed through the lens of quantified nuclear models. In Fig.~\ref{fig:APV-polariz-48vs208} we compare predictions of theoretical models for
  $\Apv$ and $\aD$ in ${}^{48}$Ca and ${}^{208}$Pb. The lower panel shows the results for $\aD$. The theoretical results line up  along a linear trend whose direction aims clearly toward the intersection of the two experimental results. Several models (except for PC-min and SV-min$^*$) are consistent with experimental data for $\aD$. The upper panel shows similar comparison  for  $\Apv$. Theoretical results exhibit again  a linear trend which, however, bypasses  the experimental intersection. The error ellipsoids show slight deviations from the linear trend, but not enough to embrace the data. The wrong direction of the average trend together with the rather narrow error ellipsoids suggest
 that a simultaneous fit of both $\Apv$ values cannot produce a consistent explanation of PREX-2 and CREX  measurements.

\paragraph{Conclusions.} In this Letter, we  critically assessed the predictions of the quantified nuclear DFT models  in the context of the recent PREX-2 and CREX measurements of  the parity-violating asymmetry $\Apv$.
Our results raise questions on: (i) the suitability of the current theory to describe  the measured  $\Apv$ values; (ii) the physical content of the correlations between  $\Apv$ and various observables/parameters; and  (iii) the
suitability of measured  $\Apv$ values  to deduce $\skin$, $J$, and $L$.
Regarding  (i) and (ii), the  EDFs employed in our study have been used  successfully to describe masses, charge radii,  giant resonances, and other nuclear properties  along the whole nuclear chart, and there is no indication that these EDFs are fundamentally wrong. Indeed, charge radii are typically described
by state-of-the-art EDFs within 0.015-0.02\,fm average deviations and masses are calculated within 1-2\,MeV. Such a  global level of agreement with experiment throughout the entire nuclear chart has not been reached by any other microscopic theoretical tool that can also address the nature of excited states. In order to explore the model dependence of the correlations, we have considered a slightly more general functional SV-ext, see SMat. Because of the extended parameter space, the correlations provided by SV-ext are slightly reduced.    Whether the physical correlations discussed in this Letter are valid for a greater class of EDFs, which go well beyond the models used here
(see, e.g., Refs.~\cite{Baldo2008,Dobaczewski2016,Navarro2018,Papakonstantinou2018,Marino2021} for the recent studies on EDF developments)
still remains to be investigated.

The results presented in Figs.~\ref{fig:APV2-trendAPV208-JL} and Fig.~\ref{fig:APV-polariz-48vs208} suggest  a tension between the $\Apv$ data and global nuclear EDFs  {\it or} that the $\Apv$ values of CREX and PREX-2  are not mutually compatible within the given experimental  errors with the current theory. This  calls for a critical search
 of limitations of current nuclear EDFs and interactions used in ab-initio calculations and/or possible  other sources of uncertainty in experiment. We also confirm the conclusion reached in Ref.~\cite{Reinhard2021}:  the significant  uncertainties, specially of PREX-2 value of $\Apv$, make it difficult to use this observable as a meaningful constraint on the isovector sector of current EDFs. 
 Until the tension between theory and experiment, or between the two measurements (see e.g., Ref.~\cite{Thiel2019} for planned MREX experiment at MESA), is resolved, one should exercise extreme caution when interpreting the new $\Apv$ measurements in the context of neutron skins or nuclear symmetry energy.

{\it Acknowledgements}.---This material is based upon work supported by the U.S.\ Department of Energy, Office of Science, Office of Nuclear Physics under award numbers DE-SC0013365 and DE-SC0018083 (NUCLEI SciDAC-4 collaboration). We thank the RRZE, the regional computing center of the university Erlangen for supplying the necessary computing resources.

\bibliography{APV}

\end{document}